\documentstyle[epsfig,12pt,pic98pro]{article}
\newcommand {\MW}      {M_{\mathrm{W}}}

\newcommand {\MH}      {M_{\mathrm{H }}}

\newcommand {\qq}      {{\rm q}\overline{\rm q}}

\newcommand {\ee}      {\mathrm{e}^+\mathrm{e}^-}
\newcommand {\WW}      {\mathrm{W}^+\mathrm{W}^-}
\newcommand {\roots}   {\sqrt{s}}
\newcommand {\Bh}     {\mathrm{B_h}}
\begin{document}
\title{ 
W Physics at LEP
}
\author{
Monica~Pepe-Altarelli,                            \\
{\em INFN - Laboratori Nazionali di Frascati}  
}
\maketitle
\baselineskip=14.5pt
\begin{abstract}
A summary of the W-boson properties measured by the four LEP collaborations
is presented here. These properties are updated to take into
account the most recent results presented at the ICHEP98 Conference. 

\end{abstract}
\baselineskip=17pt
\section{WW cross sections and W branching fractions}
After the period of running at the Z, the centre-of-mass energy of LEP has
been progressively increased from 161~GeV, i.e., 
just above the W pair production threshold
to the current centre-of-mass energy of 189~GeV.
Each LEP experiment has collected a luminosity of approximately
10~pb$^{-1}$ at $\roots=161$~GeV, 10~pb$^{-1}$ at $\roots=172$~GeV
and 55~pb$^{-1}$ at $\roots=183$~GeV. The current run at $\roots=189$~GeV
is expected to yield a luminosity of $\sim150$~pb~$^{-1}$; preliminary
results on the $\WW$ cross section at $\roots=189$~GeV based on the
data analysed for ICHEP98 (i.e., about 36~pb~$^{-1}$ per experiment) are also
reported.
  
\begin{figure}[htbp]
\begin{center}
\mbox{
\epsfig{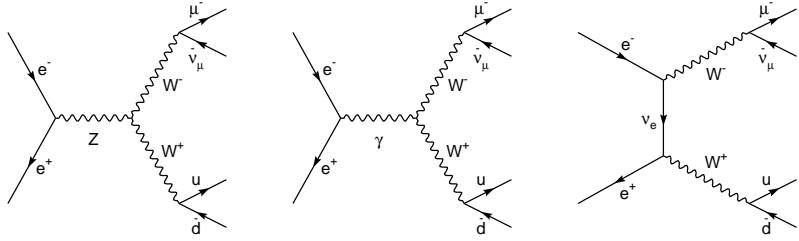}} 
\caption{\it
Signal diagrams for $\ee\to\mu\nu_{\mu}\rm{u\bar{d}}$.
\label{feyn_sig} }
\end{center}
\end{figure}

To lowest order, three Feynman diagrams contribute to W pair production
at LEP~II, the s-channel Z and $\gamma$ exchange and the t-channel $\nu_e$
exchange (the so-called CC03 diagrams displayed in
Figure~\ref{feyn_sig} for the $\mu\nu_{\mu}\rm{u\bar{d}}$ final state). 
The s-channel diagrams arise 
as a consequence of the trilinear gauge-boson vertices $\gamma\WW$ and Z$\WW$.
The process which is experimentally relevant is
$\ee\to\WW\to\mathrm{f_1\bar{f}_2f_3\bar{f}_4}$.
Many more diagrams contribute to four-fermion production.
Those with the same final states as for $\WW$ production
interfere with the signal processes. Therefore, to obtain 
the $\WW$ cross sections corresponding to the three CC03 diagrams, 
the measurements have to be corrected for
four-fermion effects.

\begin{figure}[t]
\begin{center}
\mbox{
\epsfig{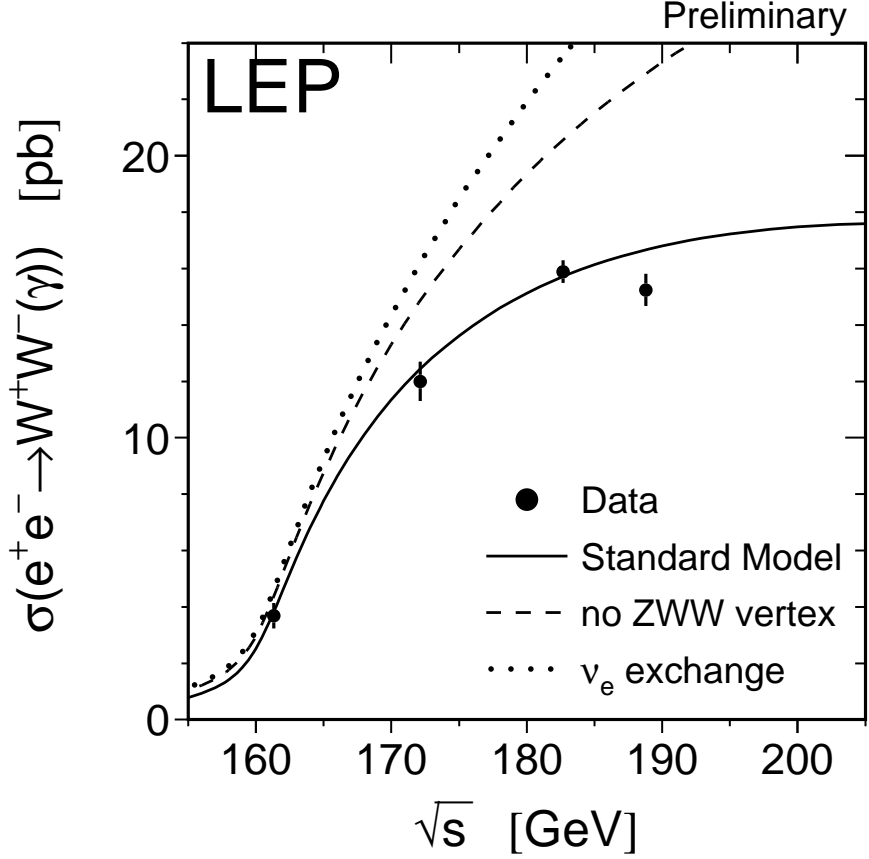}} 
\caption{\it
      W-pair cross section as a function of $\roots$. 
The data points are the combined LEP cross sections. The curves show
the Standard Model prediction (full line), the calculated cross section
if no ZWW vertex existed (dashed line) and if both ZWW and $\gamma$WW vertices
did not exist (dotted line).
    \label{cross_en} }
\end{center}
\end{figure}

Table~\ref{W-cross} gives the values of the cross sections for
the four experiments combined at the different centre-of-mass energies~\cite{mw_combi, karlen, malgeri, LEPEW}.
\begin{table}[b]
\centering
\caption{ \it $\WW$ cross sections for the four LEP experiments combined
for the various centre-of-mass energies. The results relative to the
data taken at $\roots=183$~GeV and 189~GeV are still preliminary.
}
\vskip 0.1 in
\begin{tabular}{|c|c|} \hline
            $\roots$~(GeV) & $\sigma_{\WW}$ (pb) \\
\hline
\hline
 161 & $3.69\pm0.45$\\
 172 & $12.05\pm0.73$\\
 183 & $15.86\pm 0.40$\\
 189 & $15.24\pm 0.57$\\
\hline
\end{tabular}
\label{W-cross}
\end{table}
Figure~\ref{cross_en} shows the measured cross sections as a function
of centre-of-mass energy together with the Standard Model (SM)
prediction (full line).
Also shown are the predictions
obtained assuming the existence of the 
t-channel $\nu_e$-exchange diagram only (dotted line) or if the ZWW vertex
did not exist (dashed line). 
While contributions from the individual Feynman graphs grow with
energy, an energy behaviour in agreement with data is only obtained when the full amplitude
is considered, due to cancellations 
which can be traced to the gauge theory relations between
fermion-gauge boson vertices and triple gauge couplings.

In Table~\ref{W-BR}, the W decay branching ratios are reported with and without assuming
lepton universality~\cite{mw_combi, karlen, malgeri}. 
Correlated errors between the various channels are taken into account
in the measurements; in particular the branching ratio for the channel W$\to\tau\nu_{\tau}$
has a correlation of -25\% with the other two leptonic channels.
\begin{table}
\centering
\caption{ \it  Summary of W branching fractions from $\WW$ cross section 
measurements up to 183~GeV centre-of-mass energy~\cite{mw_combi, karlen, malgeri}.
}
\vskip 0.1 in
\begin{tabular}{|l|c|c|c|c|} \hline
          & W$\to \rm{e}\nu$ & W$\to \mu\nu$    &W$\to \tau\nu$      & W$\to$hadrons\\
Experiment&     (\%)         &   (\%)           &  (\%)              &  (\%)        \\
\hline
ALEPH     &$11.2\pm0.8\pm0.3$&$9.9\pm0.8\pm0.2 $ & $9.7\pm1.0\pm0.3 $ & $69.0\pm1.2\pm0.6$\\
DELPHI    &$9.9\pm1.1\pm0.5 $&$11.4\pm1.1\pm0.5$ & $11.2\pm1.7\pm0.7$ & $67.5\pm1.5\pm0.9$\\
L3        &$10.5\pm0.9\pm0.2$&$10.2\pm0.9\pm0.2$ & $9.0\pm1.2\pm0.3 $ & $70.1\pm1.3\pm0.4$\\
OPAL      &$11.7\pm0.9\pm0.3$&$10.1\pm0.8\pm0.3$ & $10.3\pm1.0\pm0.3$ & $67.9\pm1.2\pm0.6$\\
\hline
LEP       & $10.92\pm0.49$   &  $10.29\pm0.47$  & $9.95\pm0.60$      & $68.79\pm0.77$  \\
\hline
LEP W$\to\ell\nu$& \multicolumn{3}{c|}{$10.40\pm0.26$}                 &                  \\
\hline
SM               & \multicolumn{3}{c|}{10.8}                           &    67.5          \\
\hline  
\end{tabular}
\label{W-BR}
\end{table}
The W hadronic branching ratio $\Bh$ can be related to the six elements of the 
CKM matrix (V$_{\rm{CKM}}$) not involving the top quark
via the formula
\begin{equation}
\frac{\Bh}{1-\Bh}= \sum_{\rm{i=u,c\,\,\,j=d,s,b}} |\rm{V_{i,j}}|^2 (1+\frac{\alpha_s}{\pi}).
\end{equation}
Since $|\rm{V_{cs}}|$ is 
rather poorly measured ($|\rm{V_{cs}}|=1.04\pm0.16$ from data on
branching ratios for $D_{e3}$ and  D lifetimes~\cite{PDG}) 
it can be determined from the above expression
by taking for the other CKM matrices the current world averages~\cite{PDG},
$\alpha_s=0.118\pm0.03$~\cite{PDG}  and not assuming unitarity of V$_{\rm{CKM}}$.
The result is~\cite{mw_combi, karlen, malgeri, obra}:
\begin{equation}
|\rm{V_{cs}}|=1.04\pm0.04.
\end{equation}
The error on this result is dominated by the statistical error on the W branching fractions.
The element $|\rm{V_{cs}}|$ can also be determined by direct flavour tagging, based, for
example, on a lifetime or D$^*$ tag. The result is however less precise in this case~\cite{karlen,obra}:
\begin{equation}
|\rm{V_{cs}}|=0.99\pm0.11.
\end{equation} 
\section{Triple gauge couplings}
The most general Lorentz invariant Lagrangian describing the triple gauge 
boson interactions has fourteen terms which reduce to five assuming
electromagnetic gauge invariance as well as P and C conservation.
Since we are interested in possible deviations from the SM, the anomalous couplings
$\Delta {g_1}^Z$, $\Delta k_Z$, $\Delta k_{\gamma}$, $\lambda_Z$ and $\lambda_\gamma$,
which are all zero in the SM at tree level, are chosen as free parameters.
The Triple Gauge boson Couplings (~TGCs) contribute via loops to observables 
which are precisely measured at LEP~I. It is therefore convenient to choose
combinations  of couplings not tightly constrained by existing LEP~I data.
This leads to the choice at LEP~II of the following parametrisation~\cite{LEP2}:
\begin{eqnarray}
 \alpha_{\rm{W}\phi} &=&  \Delta {g_1}^Z \rm{cos^2 \theta_W} \\
 \alpha_{\rm{W}}     &=&  \lambda_\gamma \\
 \alpha_{\rm{B}\phi} &=&  \Delta\kappa_\gamma - \Delta {g_1}^Z \rm{cos^2 \theta_W}
\end{eqnarray}
with the constraints $\lambda_{\rm{Z}}=\lambda_\gamma$ and 
$\Delta\kappa_{\rm Z}=-\Delta\kappa_\gamma\rm{tan}^2\theta_{\rm W} +\Delta {g_1}^Z$
if SU(2)$\otimes$SU(1) gauge invariance is  also required. The $\alpha$-parameters are
all zero in the SM at tree level.

Anomalous couplings increase the $\WW$ cross section and
change the angular distribution of the produced W bosons and of  
their decay products. The strongest information is provided by the semileptonic
$\qq\ell\nu$ decay for which the charge assignment is unambiguous. On the contrary,
in the case of the hadronic channel, without quark charge or flavour tagging,
the fermion and anti-fermion cannot be distinguished.
The differential cross sections in the production and decay angles are written in terms
of the W helicity amplitudes, which are, in turn, well defined functions of the TGCs. 

\begin{figure}[htb]
\begin{center}
\mbox{
\epsfig{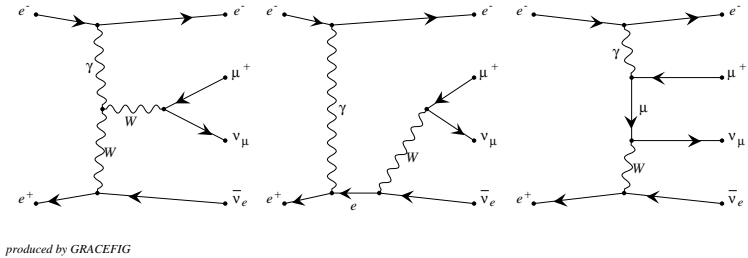}} 
\caption{\it
Dominant diagrams for the $\ee\to\rm(e)\nu\mu\nu$ 
final state.
    \label{feyn_1w} }
\end{center}
\end{figure}

\begin{figure}[htb]
\begin{center}
\mbox{
\epsfig{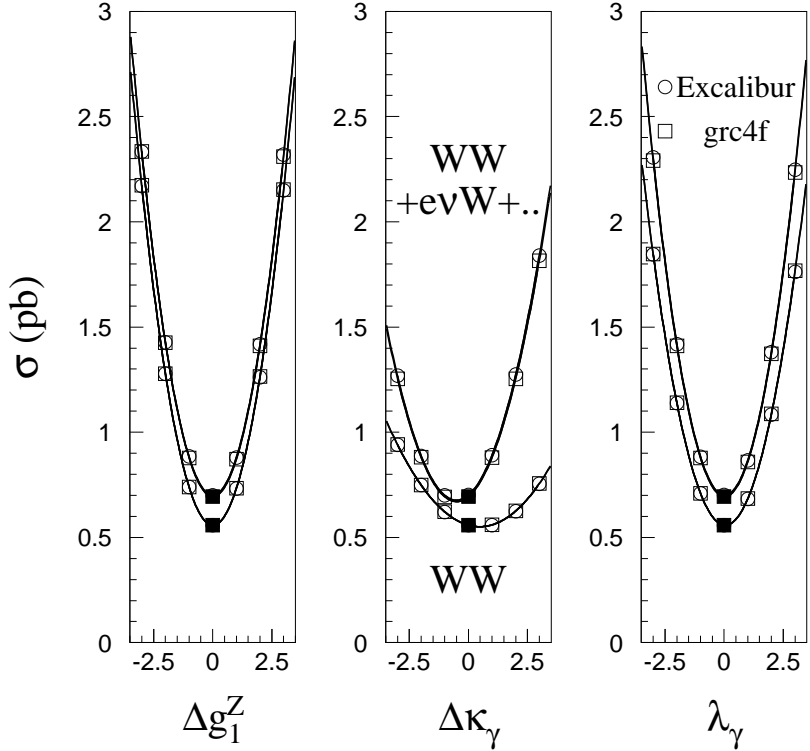}} 
\caption{\it
Sensitivity of the cross sections to $\Delta {g_1}^Z$, $\Delta\kappa_\gamma$ and 
$\lambda_\gamma$  with and without inclusion of the single W processes.
    \label{hagiwara2} }
\end{center}  

\end{figure}

Additional sensitivity to the WW$\gamma$ couplings is provided by single photon
and especially single W events.
The dominant diagrams for single W production in the
e$\nu\mu\nu$ final state are shown in Figure~\ref{feyn_1w},
the first diagram being the one providing sensitivity to the WW$\gamma$ vertex.
Since, typically, the electron goes down the beam pipe,
the signatures are a single energetic lepton, or two acoplanar jets and large
missing energy.
Figure~\ref{hagiwara2} shows the gain in sensitivity on $\Delta\kappa_\gamma$
(and therefore on  $\alpha_{\rm{B}\phi}$) obtained when the ``standard'' WW analysis
is combined with the single W analysis~\cite{tanaka}.

\begin{figure}[htb]
\begin{center}
\mbox{
\epsfig{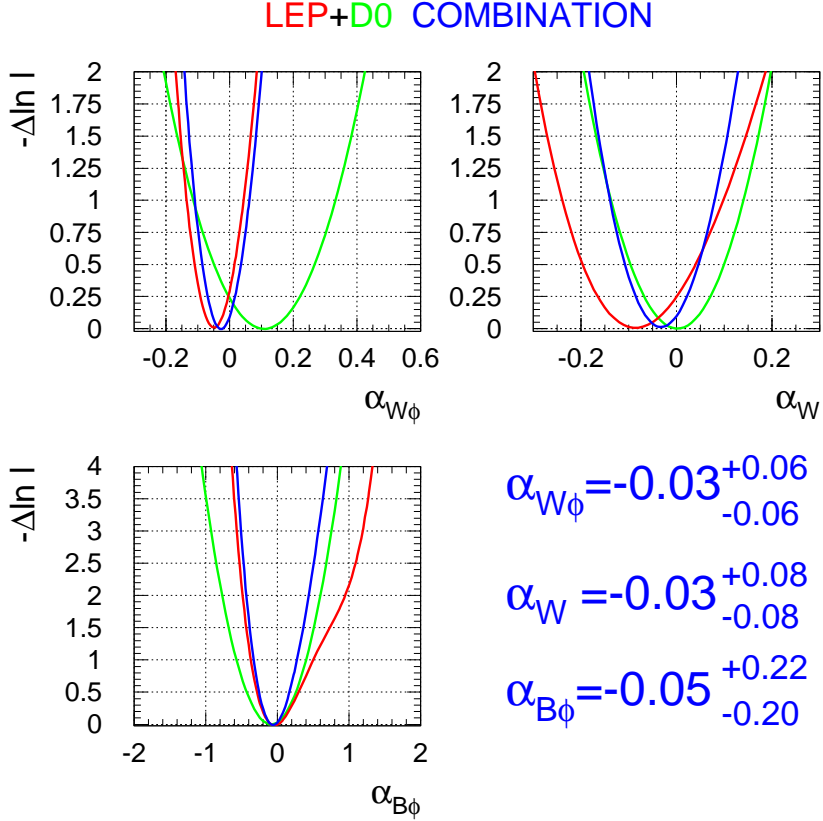}} 
\caption{\it
Results for the three $\alpha$ couplings combining LEP (dark grey line), 
D0 (light grey line)
and LEP+D0 (black line).
    \label{tgc_all} }
\end{center}
\end{figure}

The results of fits to
$\alpha_{\rm{W}\phi}$, $\alpha_{\rm{W}}$ and $\alpha_{\rm{B}\phi}$ are shown in 
Figure~\ref{tgc_all}, assuming in each case that the other two anomalous couplings
are zero. The combination is performed by adding the log-likelihood curves
supplied by the LEP and D0~\cite{TGC_comb} experiments. 
The one standard deviation and 95\% confidence limits are taken
as the parameter values where $-\Delta\rm{log\,L}=0.5$ and 1.92, respectively.
No discrepancy with the SM is observed, however the accuracy on the determination of the
$\alpha$ parameter is  rather poor (in the SM, from radiative corrections,
they are expected to be of order $10^{-2}$, $10^{-3}$). 
\section {Measurement of the W mass}
The data collected at centre-of-mass energy of $161$~GeV, i.e., just above
the W pair production threshold, were used to obtain $\MW$
by comparing the measured cross section with a theoretical calculation 
which has $\MW$ as a free parameter.
In fact, for $\roots\simeq 2\MW$, the value of the cross section is very sensitive
to the W mass and the precise value of 161~GeV was chosen as the best
compromise between sensitivity to $\MW$ and statistics. Despite the fact
that one relies on a theoretical calculation based on the SM to derive $\MW$, the
model dependence is small since at threshold the cross section is dominated
by the well established $\nu_{\rm{e}}$ t-channel exchange diagram.
From the LEP combined cross section at 161~GeV $\sigma_{\rm{WW}}=3.69\pm0.45$~pb,
the following result for $\MW$ is obtained:
\begin{equation}
\MW=80.40\pm0.22 \,\rm{GeV}.
\end{equation}

At higher energies $\MW$ is obtained by direct reconstruction of the jet-jet
invariant masses from the channels:
\begin{eqnarray}
\WW \to \qq\qq     \, & (\sim45\%\,\rm{of\,cases}\,) \nonumber\\
\WW \to \qq\ell\nu \,& (\sim44\%\,\rm{of\,cases}\,). \nonumber
\end{eqnarray}
These topologies are selected with high efficiency and low background.
Efficiency and purity for 
the fully hadronic events are
approximately $85\%$. The semileptonic events are selected with typically 85\% ( 60\%)
efficiency and 95\% ( 85\%) purity
for e and $\mu$ ($\tau$).
The purely leptonic channel has not been used so far due to the lack of sufficient
constraints on the kinematics of the event, as it contains at least two neutrinos.

A kinematic fit requiring energy and momentum conservation 
is used to improve the invariant mass resolution. An additional constraint based
on the equality of the two W masses in an event is also frequently used.
In the four-jet case, three kinematic fits are performed
for the three possible jet-jet pairings and the resulting fit probabilities
are used to discard, in general, at least one of the three combinations.

The invariant mass distribution has a Breit-Wigner shape which is distorted
by several effects such as initial state radiation, detector resolution, misassignement
of particles between the two W bosons, background, analysis biases, etc, which can
only be evaluated by extensive Monte Carlo simulations.
What is generally done to extract $\MW$ is to compare the measured invariant
mass spectra with the corresponding distributions from
simulated experiments based on different input W masses.
To avoid generating large Monte Carlo samples at many different
masses, starting from a few reference input W masses,
the mass spectra corresponding to other choices of $\MW$ are
obtained by reweighting each event in the reference Monte Carlo
by the ratio of cross sections calculated with the new and the reference W mass.

Table~\ref{mw_tab} shows a summary of the W mass measurement by direct
reconstruction (at 172 and 183~GeV) for the four LEP experiments
in the various decay modes. The combined four-experiment W mass is~\cite{mw_combi,karlen,helenka,thomson}:
\begin{eqnarray}
\MW^{\rm{lept}}(172-183)&=& 80.31\pm0.10_{\rm{stat}}\pm0.03_{\rm{syst}}\pm0.025_{\rm {LEP}} \nonumber\\
\MW^{\rm{had}} (172-183) &=& 80.39\pm0.093_{\rm{stat}}\pm0.05_{\rm{syst}}\pm0.09_{\rm{FSI}}\pm0.025_{\rm {LEP}} \nonumber\\
\MW(172-183)           &=& 80.36\pm0.08\pm0.05_{\rm{FSI}}\pm0.025_{\rm {LEP}}\,. \nonumber
\end{eqnarray}
\begin{table}
\centering
\caption{ \it Summary of W mass measurements by direct reconstruction
(~i.e. using data at $\roots$=172 and 183~GeV) for the four LEP experiments in the various
decay channels~\cite{mw_combi,karlen,helenka,thomson}. The errors reported include both statistical
and systematic uncertainties.
}
\small
\vskip 0.1 in
\begin{tabular}{|l|c|c|c|} \hline
             & Semileptonic    & Hadronic  & Combined \\
  Exp.  &  $\MW$~(GeV)    & $\MW$~(GeV)& $\MW$~(GeV) \\
\hline
 ALEPH  & $80.34\pm0.18$ & $80.53\pm0.18$ & $80.44\pm0.13$ \\
 DELPHI & $80.50\pm0.24$ & $80.01\pm0.22$ & $80.24\pm0.17$ \\
 L3     & $80.09\pm0.24$ & $80.59\pm0.23$ & $80.40\pm0.18$ \\
 OPAL   & $80.29\pm0.19$ & $80.40\pm0.24$ & $80.34\pm0.15$ \\
\hline
Combined& $80.31\pm0.11$ & $80.39\pm0.14$ & $80.36\pm0.09$ \\
\hline
\end{tabular}
\label{mw_tab}
\end{table}
\normalsize
The semileptonic and hadronic channels have comparable branching ratios and selection
efficiencies and give comparable mass resolution. However, the hadronic mode 
has an additional systematic error of 90~MeV
associated to final state interaction effects (FSI)
which represents the largest source of systematic uncertainty,
as shown in Table~\ref{tab_syst}. 
\begin{table}
\centering
\caption{ \it
``Typical'' systematic uncertainties for the semileptonic and
hadronic channels. The upper (lower) part of the Table gives
those errors which are not correlated (correlated) between experiments.
}
\vskip 0.1 in
\begin{tabular}{|l|c|c|} \hline
  Systematic        & Semileptonic & Hadronic \\
  source            &   \multicolumn{2}{c|}{$\delta\MW$~(MeV)}\\
\hline
Detector calibration     & 40  & 30 \\
QCD background           &     & 10 \\
MC statistics            & 10  & 10 \\
\hline
Hadronisation            & 25  & 30 \\
ISR                      & 15  & 15 \\
Beam energy              & 25  & 25 \\
Final State Interactions &  -  & 90 \\
\hline
 TOTAL                   &$\sim$~60& 110\\ 
\hline
\end{tabular}
\label{tab_syst}
\end{table}

Final state interactions may arise since the separation of the W decay vertices at LEP~II
is $\sim0.1$~fm, a distance small with respect to the typical hadronisation
scale of $\sim$1~fm (or, in other words, $\Gamma_{\rm{W}}\sim10\Lambda_{\rm{QCD}}$).
As a result, interconnection phenomena may obscure the separate identities
of the two W bosons distorting the mass determination in the hadronic channel. 
These interconnection effects can be associated to colour fields
stretched between quark lines from different W bosons (``Colour Reconnection'') or to interference
between identical bosons close in phase space, but produced by different W decays
(``Bose-Einstein correlations'').

The best test of colour reconnection is realized by comparing mean values of charged particle
multiplicity and event-shape distributions in the fully hadronic and semileptonic modes
since many systematic effects cancel in the difference.
The distributions for the $\qq\qq$ mode should be equal to twice the $\qq\ell\nu$ mode
after removing the final state lepton or its decay products.
The following result on the average charged multiplicity
difference between $\qq\qq$ and $\qq\ell\nu$ is obtained by combining the four LEP experiments~\cite{karlen,watson}:
\begin{equation}
\Delta n_{\rm{ch}}\equiv\langle{ n_{\rm{ch}}}^{\qq\qq}\rangle-
2\langle{n_{\rm{ch}}}^{\qq\ell\nu}\rangle=0.20\pm0.50\,.
\end{equation}
At the current level of statistical precision, no evidence for colour reconnection effect
is found in the observables studied. Most models, such as Sj\"{o}strand-Khoze~\cite{khoze}, 
ARIADNE~\cite{ARIADNE}, HERWIG~\cite{HERWIG}, are consistent with the data and predict
shifts for $\MW$ smaller than $\sim$50~MeV.
The Ellis-Geiger model~\cite{ellis} has not been used to estimate the systematic error since,
in its current implementation, does not reproduce a variety of the measured event shapes.
 
The simplest method to analyse Bose-Einstein correlations is to
measure the ratio of like-sign to unlike-sign pion pairs as a function of $Q^2=(p_1-p_2)^2$ where $p_1$
and $p_2$ are the particle four-momenta. Contribution of pion pairs originating from the same W
are subtracted statistically, using the distribution obtained from semileptonic events.
Bose-Einstein correlations are observed in hadronic and semileptonic W decays. However,
at the present level of statistics, there is no experimental evidence for
Bose-Einstein correlations for pairs originating from different W bosons~\cite{karlen, moller}.
Phenomenological studies indicate that this effect could introduce a shift to $\MW$   
smaller than $\sim$50~MeV.
On the basis of the present experimental and phenomenological results,
the error of 90~MeV which is currently assigned as common systematic uncertainty
between the LEP experiments due to final state interactions (Bose-Einstein and Colour Reconnection
effects) is probably a rather conservative estimate.  

\begin{figure}[htb]
\begin{center}
\mbox{
\epsfig{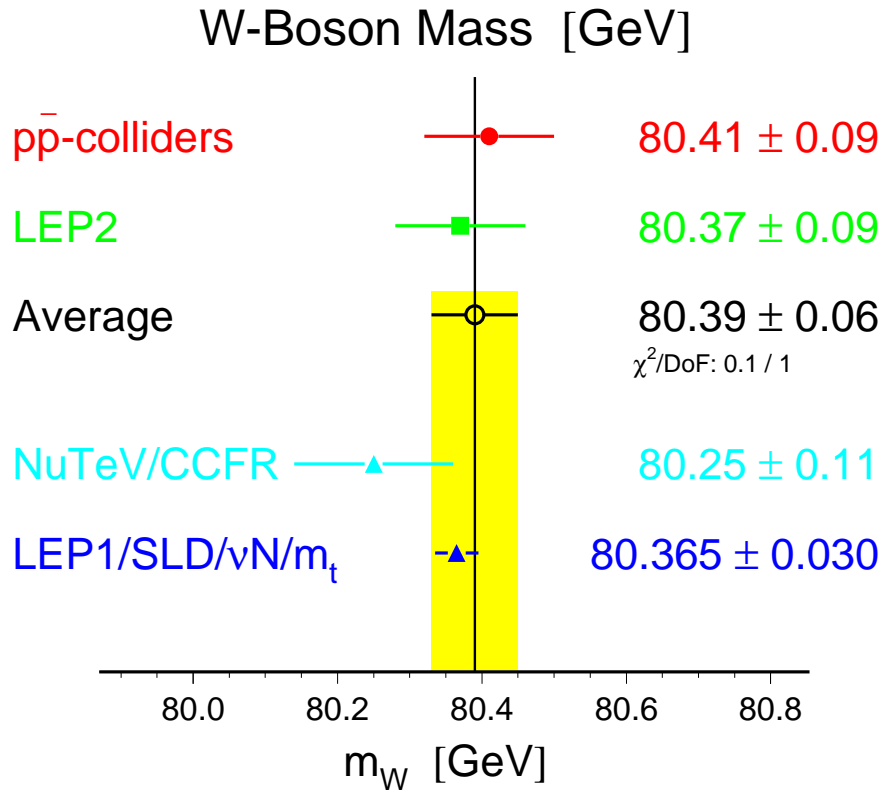}} 
\caption{\it
Summary of $\MW$ measurements from LEP, the TEVATRON experiments and $\nu$N scattering as well
as the indirect determination derived from all other electroweak data.    \label{mw_all} }
\end{center}
\end{figure}

Since the systematic uncertainties of the direct reconstruction technique and
threshold method are largely independent, the two measurements can be combined,
yielding the following result:
\begin{equation}
\MW=80.37\pm0.09\,\rm{GeV}.
\end{equation} 
Figure~\ref{mw_all} shows a summary of direct determinations of $\MW$ from LEP and the TEVATRON,
as well as the indirect estimates from $\nu$N scattering 
and from radiative corrections using all 
electroweak data~\cite{karlen,thomson}. Direct $\MW$ measurements provide a precision
which is approaching the one obtained by
radiative corrections, allowing a further important test of the SM.
The goal of measuring by the end of the LEP~II programme in the year 2000
the W mass with a precision of $\sim$30-40~MeV seems to be in reach.
Since $\MW$ is an observable which is  sensitive to $\MH$, this measurement
will allow to put additional constraints on the Higgs mass. The real break-through
would be of course the discovery of the Higgs in the $\sim$10~GeV mass window
which is still accessible to LEP~II.

\section{Acknowledgements}
I would like to express my sincere gratitude to the organisers, to Franco Fabbri in particular,
for their invitation and for the nice organisation of the conference
in such a beautiful setting.

%
\end{document}